\documentclass[preprint,prd,noshowpacs,nofootinbib]{revtex4}
\usepackage{bm}
\usepackage{enumitem}   
\usepackage{latexsym}
\usepackage[dvipsnames]{xcolor}
\usepackage{dcolumn}
\usepackage{amsmath,amsfonts,amssymb}
\usepackage{graphicx,float}
\usepackage{appendix}
\usepackage[hyperfootnotes=true]{hyperref}
\usepackage{mciteplus}
\usepackage{subcaption}
\usepackage{caption}
\usepackage{bbold}
\usepackage{slashed}
\usepackage{mathbbol}
\DeclareSymbolFontAlphabet{\amsmathbb}{AMSb}%
\usepackage{amsthm}
\usepackage{hyperref}
\usepackage{color}
\usepackage{mathrsfs}
\usepackage{setspace}

\DeclareMathAlphabet{\mathpzc}{OT1}{pzc}{m}{it}
\usepackage{calligra}
\DeclareMathAlphabet{\mathcalligra}{T1}{calligra}{m}{n}
\DeclareFontShape{T1}{calligra}{m}{n}{<->s*[2.2]callig15}{}

\def\be {\begin{equation}}
\def\ee {\end{equation}}
\def\bea {\begin{eqnarray}}
\def\eea {\end{eqnarray}}
\def\bc {\begin{center}}
\def\ec {\end{center}}
\def\bfg {\begin{figure}}
\def\efg {\end{figure}}
\def\bi {\begin{itemize}}
\def\ei {\end{itemize}}

\def\beq{\begin{equation}}
\def\eeq{\end{equation}}
\def\br{\begin{eqnarray}}
\def\er{\end{eqnarray}}
\newcommand{\eel}[1] {\label{#1}\end{equation}}

\newcommand{\bdm}{\begin{displaymath}}
\newcommand{\edm}{\end{displaymath}}

\begin{document}

\title{Could GUP Act as a Model for the ER=EPR Conjecture?}
\author{Ahmed Farag Ali $^\nabla$$^{\triangle}$}
\email[email: ]{aali29@essex.edu}
\affiliation{{$^{\nabla}$Essex County College, 303 University Ave, Newark, NJ 07102, United States }}
\affiliation{ $^{\triangle}$ Dept. of Physics, Benha University, Benha 13518, Egypt}

\begin{abstract}
\par\noindent
Einstein, Podolsky, and Rosen (EPR) proposed, via a thought experiment, that the uncertainty principle might not provide a complete description of reality. We propose that the linear generalized uncertainty principle (GUP) may resolve the EPR paradox by demonstrating vanishing  uncertainty at the minimal measurable length. This may shed light on the completeness of quantum mechanics which leads us to propose an equivalency between the linear GUP and the Bekenstein  bound, a bound that prescribes the maximum amount of information needed to completely describe a physical system up to quantum level. This equivalency is verified through explaining the Hydrogen's atom/nuclei radii as well as the value of the cosmological constant. In a recent published study, we verified that the Einstein-Rosen (ER) bridge originates from the minimal length or GUP. Considering these findings together, we propose that linear GUP could function as a  model for the ER=EPR conjecture.

%\begin{center}
% ``Movement is constant'' THOTH\\
%    “Truth is ever to be found in simplicity”  NEWTON.
%\end{center} 

\end{abstract}

\maketitle

\tableofcontents

\section{introduction}
\par\noindent  In 1935, Einstein, Podolsky, and Rosen (EPR) presented a seminal critique of quantum mechanics, proposing a framework where predictions must correlate with observable realities to deem a theory complete. They argued that properties predicted without direct interaction are inherently real and that a theory's inability to explain such properties indicates its incompleteness. Central to this discourse is the uncertainty principle, encapsulated by the commutation relation:
\begin{equation}
    [x_i, p_j] = i \delta_{ij} \hbar,
\end{equation}
where $x$ signifies the position operator, $p$ the momentum operator, and $\hbar$ the reduced Planck's constant. This principle suggests the impossibility of simultaneously determining a particle's precise momentum and position, attributed to the observational disturbance. Contrary to this, the EPR paradox challenges the quantum mechanical framework by asserting the possibility of determining a particle's momentum or position without direct interaction. Through a carefully constructed thought experiment involving two particles, A and B, that interact and then diverge, EPR highlighted the limitations imposed by the uncertainty principle. Despite quantum mechanics' restriction on the simultaneous knowledge of momentum and position, it allows for the collective momentum measurement of the particle pair. EPR posited that the momentum of particle B could be deduced from the momentum of particle A, using the conservation of momentum principle, without direct observation of B. This deduction underscores the tangible, determinable nature of particle B's momentum. Further, EPR contended that the mere observation of particle A is sufficient to infer both the velocity and position of particle B without any interference, challenging the quantum mechanical stance that definitive properties cannot coexist as determinate values absent direct observation. Therefore, the EPR paradox not only questions the completeness of quantum mechanics but also prompts a reconsideration of the theory's ability to articulate the complexities of physical reality, especially regarding the determinability of properties without direct observation. This paradox has ignited a critical philosophical debate within quantum physics, marking a significant moment in the discourse on the interpretation of reality through the quantum mechanical lens. A very useful review on this issue can be found in \cite{kumar2008quantum}. EPR conclusion in their own words are as follows
\begin{quote}
\emph{``Previously we proved that either (1) the quantum-mechanical description of reality given by the wave function is not complete or (2) when the operators corresponding to two physical quantities do not commute the two quantities cannot have simultaneous reality. Starting then with the assumption that the wave function
does give a complete description of the physical reality, we arrived at the conclusion that two physical quantities, with noncommuting operators, can have simultaneous reality. Thus the negation of (1) leads to the negation of the only
other alternative (2). We are thus forced to conclude that the quantum-mechanical description of physical reality given by wave functions
is not complete.''Einstein, Podolsky, and Rosen said} \cite{Einstein:1935rr}.
\end{quote}
EPR suggested a local hidden variable theory to resolve their paradox. Bohm investigated that possibility  \cite{Bohm:1951xw} through the quantum potential and pilot-wave approach. However, Bell's inequalities \cite{bell1964einstein}, developed later, interpreted the EPR paradox differently, suggesting non-local features in quantum mechanics. In essence, Bell's theory did not resolve the paradox but offered a new interpretation based on non-locality. It implies that any hidden-variable theory that might underlie quantum mechanics has to be non-local. These features have been subsequently verified through entanglement experiments \cite{aspect1982experimental}. In recent times, an intriguing proposition has been put forth suggesting a potential resolution to the EPR paradox. This perspective introduces the concept of a wormhole, or Einstein-Rosen bridge, as the connecting thread in the EPR paradox. This theoretical proposition, known as the ER=EPR conjecture, posits that entangled particles are connected by a wormhole. It's a groundbreaking idea that aims to reconcile the seemingly incompatible aspects of quantum mechanics and general relativity \cite{Maldacena:2013xja}.\\

\noindent
Given that physical properties can exist independently of direct observation, it follows that these properties may be correlated with a minimal measurable length. Consequently, we propose that a modified uncertainty principle incorporating a minimal length could provide a framework for a physical state where traditionally incompatible quantities, such as position and momentum, can possess simultaneous reality. We investigate modified theories of the uncertainty principle, collectively referred to as the generalized uncertainty principle (GUP) \cite{Snyder:1946qz,Amati:1988tn,Garay:1994en,Scardigli:1999jh,Brau:1999uv,Kempf:1994su,Maggiore:1993rv,Capozziello:1999wx,Todorinov:2018arx,Ali:2009zq}. Various quantum gravity theories, such as string theory, loop quantum gravity, and quantum geometry, propose a generalized version of the uncertainty principle (GUP), which suggests the presence of a minimum measurable length. In earlier research, multiple forms of the GUP have been introduced, including both non-relativistic and relativistic types \cite{Amati:1988tn,Garay:1994en,Scardigli:1999jh,Brau:1999uv,Kempf:1994su,Maggiore:1993rv,Capozziello:1999wx,Todorinov:2018arx,Ali:2009zq}. The GUP's phenomenological and experimental implications have been examined across low and high-energy contexts, encompassing atomic systems \cite{Ali:2011fa, Das:2008kaa}, quantum optical systems \cite{Pikovski:2011zk}, gravitational bar detectors \cite{Marin:2013pga}, gravitational decoherence \cite{Petruzziello:2020wkd}, composite particles \cite{Kumar:2019bnd}, astrophysical systems \cite{Moradpour:2019wpj}, condensed matter systems \cite{Iorio:2017vtw}, and macroscopic harmonic oscillators \cite{Bawaj:2014cda}. Comprehensive reviews discussing the GUP, its phenomenology, and its experimental consequences can be found in \cite{Addazi:2021xuf,Hossenfelder:2012jw}. 
Our findings suggest that the GUP algebra may indicate a disappearing uncertainty between position and momentum at the minimal measurable length scale, leading to a simultaneous reality for both position and momentum. Therefore, the Generalized Uncertainty Principle (GUP) algebra has the potential to serve as a framework for resolving the EPR argument and could enhance our understanding of quantum entanglement within minimal-length of measurement scenarios.
Therefore, GUP offers a mathematical solution for the EPR paradox that could potentially explain the origin of non-locality in quantum mechanics and clarify how position and momentum can coexist simultaneously. In a recent collaborative work \cite{ERminimal}, we derived an Einstein-Rosen (ER) bridge using a string T-duality corrected pair of regular black holes, thereby associating it with the Generalized Uncertainty Principle (GUP) as a representation of the minimal length. This geometric model could potentially enhance our comprehension of quantum entanglement for particle/antiparticle pairs. Our findings demonstrated the feasibility of an ER bridge, where the horizon area corresponds to the Bekenstein minimal area bound in extreme mass configurations, which may hold implications for gravitational self-completeness and quantum mechanical mass limits. Taking into account the observed relationships between GUP and EPR on one hand, and ER on the other, we conclude that GUP may serve as a model manifesting the ER=EPR conjecture \cite{Maldacena:2013xja}.\\

\noindent
The structure of this paper is as follows: Section (\ref{sec:QGUP}) explores the quadratic GUP using the EPR argument. In Section (\ref{sec:LGUP}), we examine the linear GUP in the context of EPR. Section (\ref{sec:BUB}) connects the linear GUP with the Bekenstein universal bound and investigates the implications of generating four fundamental physical states characterized by positive, negative, zero, and maximum momentum. In Section (\ref{sec:EAL}), we establish a link between the linear GUP and von Neumann entropy, which gives rise to the informational origin of the momentum concept and sets a conceptual connection between a comprehensive description of physical reality implied by the linear GUP and the preservation of information as suggested by the von Neumann entropy correction. In Section (\ref{sec:sym}), we delve into the nature of time and the symmetry implied by our hypothesis. Lastly, Section (\ref{sec:con}) presents our conclusions.

\section{Quadratic GUP}\label{sec:QGUP}
\par\noindent
The GUP motivated by string theory takes the following form:
\begin{eqnarray}
\label{gupquadratic}
\Delta x\Delta p &\geq& \frac{\hbar}{2}(1+\beta\Delta p^2), \nonumber\\
\left[x_i,p_j\right]&=&i\hbar\left[\delta_{ij}+\beta \delta_{ij}p^2+2\beta p_i p_j\right]
\end{eqnarray}
where $\beta=\beta_0 \ell_p^2/\hbar^2$, $\beta_0$ is a dimensionless constant, and $\ell_p=1.6162\times 10^{-35}\text{m}$ is the Planck length. In 1926, Dirac realized that the commutator between any two variables in quantum mechanics is \emph{isomorphic} to the Poisson bracket \cite{Dirac:1926jz}. The isomorphic Poisson bracket to the quadratic GUP commutator has been studied in several physical systems in \cite{Chang:2001bm,Casadio:2020rsj,Mignemi:2014jya,Ghosh:2013qra,Guo:2015ldd,Scardigli:2014qka,Tkachuk:2012gyq} and is given by,
\begin{eqnarray}
 \{x_i,p_j\}= \left[\delta_{ij}+\beta \delta_{ij}p^2+2\beta p_i p_j\right]\text{where}~~i,j=1,2,3\label{QPoisson}
\end{eqnarray}
where $x_i$ and $p_j$ are now c-numbers. If we look at the quadratic GUP model in equation (\ref{QPoisson}),  we see in a one-dimensional case that the Poisson bracket can equal zero if we allow the right-hand side of the equation 
(\ref{QPoisson}) to be zero as follows
\begin{eqnarray}
1+3~\beta p^2=0
\end{eqnarray}
which has an imaginary solution as follows
\begin{eqnarray}
&&p= \pm~ i\sqrt{\frac{1}{3~\beta}} 
\nonumber \\
 &&\text{minimal length of measurement}= i \sqrt{3}~\beta_0 \ell_p \label{minimallength1}
\label{solution1}
\end{eqnarray}
 Let us generalize the solution of equation (\ref{solution1}) in 3 dimensions. We write equation (\ref{QPoisson}) as follows:
\begin{eqnarray}
 &&\{x_i,p_i\}= \left[3+5~\beta ~p^2\right]\text{where}~~i=j~\text{and}~ i=1,2,3  \label{QPoisson1}\\ 
 &&\{x_i,p_j\}= 2~\beta~ p_i~p_j~\text{where}~~i\neq j~ \text{and}~i,j=1,2,3
 \label{QPoisson2}
\end{eqnarray}
We used in Eq. (\ref{QPoisson1}) the property $\delta_{ii}=3$ in 3 dimensions.  We find that the uncertainty between the position and momentum \emph{of the same dimension} in equation (\ref{QPoisson1}) vanishes when 
\begin{eqnarray}
    &&p^2= \frac{-3}{5 \beta}
\end{eqnarray}
While this solution suggests the absence of uncertainty between the position and momentum of the same dimension as represented by equation (\ref{QPoisson1}) (i.e $i=j$), it still implies uncertainty between the position and momentum of different dimensions, as indicated by equation (\ref{QPoisson2}) (i.e $i\neq j$). Based on the isomorphism between the Poisson bracket and quantum commutator, the solution in equation \ref{solution1} establishes a potential state with a vanishing commutator, thus allowing for a simultaneous reality of $x$ and $p$. However, we find that for $x$ and $p$ to possess a simultaneous reality (both as real measurable values for the same physical state), the momentum must be imaginary, which is inconsistent. This imaginary solution can become real if we permit a negative value for $\beta$, but doing so would create a paradox with the conceptual foundation of the quadratic GUP as a model that predicts a minimum measurable length. Allowing for a negative $\beta$ would eliminate the minimal length concept in the quadratic GUP described by equation (\ref{gupquadratic}). As such, the quadratic GUP may possess a loophole that allows for a comprehensive representation of quantum mechanics, and the EPR argument remains valid.
\section{linear GUP}
\label{sec:LGUP}
\par\noindent
Let's consider another model, the linear GUP, which is inspired by doubly special relativity (DSR) proposed by Magueijo and Smolin \cite{Magueijo:2001cr}. DSR posits the existence of an invariant length/energy scale, in addition to the invariance of the speed of light. The corresponding uncertainty principle for doubly special relativity was first introduced in \cite{Cortes:2004qn} and later examined in \cite{Ali:2009zq,Das:2010zf}. Furthermore, when the linear GUP was analyzed in conjunction with the Schrödinger equation, Klein-Gordon equation, and Dirac equation, both a discrete representation of space and a discrete representation of energy emerged from the same wavefunction solutions \cite{Ali:2009zq,Das:2010zf}. These discrete outcomes have been obtained in both weak gravity \cite{Deb:2016psq} and strong gravity cases \cite{Das:2020ujn}. The linear GUP is expressed as follows:
\begin{eqnarray}
\label{gup}
[x_i, p_j] = i \hbar\hspace{-0.5ex} \left[  \delta_{ij}\hspace{-0.5ex}
- \hspace{-0.5ex} \alpha\hspace{-0.5ex}  \left( p \delta_{ij} +
\frac{p_i p_j}{p} \right)
 \right],~~~
\end{eqnarray}
where $\alpha=\alpha_0 l_p/\hbar$, and $\alpha_0$ is a dimensionless constant. There could be more corrections related to different forms of GUP that can be added in equation (\ref{gup}), but in this letter, we are interested to see the pure effect of the linear term in momentum in GUP and whether it can imply real solution at which the commutation relation vanishes.  The corresponding Poisson bracket of linear GUP takes the form \cite{Ali:2011ap}
\begin{eqnarray}
\label{gupPoisson}
\{x_i, p_j\} = \hspace{-0.5ex} \left[  \delta_{ij}\hspace{-0.5ex}
- \hspace{-0.5ex} \alpha\hspace{-0.5ex}  \left( p \delta_{ij} +
\frac{p_i p_j}{p} \right)
\right]~~~\text{where}~~i,j=1,2,3 \label{Pgup}
\end{eqnarray}
where $x_i$ and $p_j$ are c-numbers in equation \ref{gupPoisson}. Let us see a case in which the Poisson bracket vanishes. This can be achieved if we allow for the right-hand side of equation \ref{gupPoisson} to be equal to zero in a one-dimensional case:
\begin{eqnarray}
 \{x, p\}=1\hspace{-0.5ex}
- 2~ \alpha ~ p  
 \hspace{-0.5ex} =0 \label{Po1}
\end{eqnarray}
 that implies a solution as follows:
\begin{eqnarray}
&& p=\frac{1}{2 \alpha} \label{real}~~~~~~~~~~\text{that yields to}: \nonumber \\
 &&\text{minimal length of measurement}= 2\alpha_0 \ell_p \label{minimallength}
\end{eqnarray}
This solution can be generalized in 3 dimensions. We write equation (\ref{gupPoisson}) as follows:

\begin{eqnarray}
&&\{x_i, p_i\} = \hspace{-0.5ex} \left[  3\hspace{-0.5ex}
- 4~ \alpha~ p
\right]~~~\text{when}~~i=j \label{gupPoisson1} \\
&&\{x_i, p_j\} = -\frac{p_i~p_j}{p} ~\text{where}~~i\neq j~ \text{and}~i,j=1,2,3 \label{gupPoisson2}
\end{eqnarray}
We used in Eq. (\ref{gupPoisson1}) the property $\delta_{ii}=3$ in 3 dimensions. We find that the uncertainty between the position and momentum \emph{of the same dimension} in equation (\ref{gupPoisson1}) vanishes when
\begin{eqnarray}
    &&p= \frac{3}{4 \alpha}
     \nonumber \\
 &&\text{minimal length of measurement}= \frac{4}{3}\alpha_0 \ell_p \label{minimallength1}
\end{eqnarray}
The solution presented in equation (\ref{gupPoisson1}) implies no uncertainty between the position and momentum of the same dimension (i.e $i=j$), while equation (\ref{gupPoisson2}) suggests uncertainty between the position and momentum of different dimensions (i.e i $\neq j$). As a result, the linear GUP provides simultaneous reality between the position and momentum of the same dimension and non-simultaneous reality between the position and momentum of different dimensions. Our analysis reveals that the quadratic GUP results in vanishing uncertainty for an "imaginary" minimal length, as shown in Eq. (\ref{solution1}). Conversely, the linear GUP implies vanishing uncertainty for a "real" minimal length, as demonstrated in Eq. (\ref{minimallength}). This preference for the linear GUP model over the quadratic GUP model is due to the lack of physical significance associated with the imaginary solution for momentum/length. Bounds for the dimensionless parameter $\alpha_0$ have been established for various physical systems \cite{Ali:2011fa, Pikovski:2011zk,Das:2021nbq}. The resulting solution is real and signifies a simultaneous reality between $x$ and $p$, owing to the isomorphism or correspondence between the quantum commutator and Poisson bracket. This simultaneous reality leads to a comprehensive description of physical reality, as outlined in the logical analysis presented in the EPR study \cite{Einstein:1935rr}. On the other hand, this simultaneous reality could be employed to explain the instantaneous and non-local correlations observed when measuring the spin of two entangled particles, as previously discussed and as evidenced by the violation of Bell's inequalities. A potential conceptual link between the minimal length implied by the linear GUP and the spin concept was proposed in \cite{Ali:2021oml}. In the following section, we propose a correspondence between the GUP and Bekenstein's universal bound.
\section{Bekenstein Universal Bound and linear GUP}
\label{sec:BUB}
\par\noindent
We demonstrate that the universal bound presented by linear GUP results in a complete describtion of physical reality. On a different note, Bekenstein established a universal bound \cite{Bekenstein:1980jp} that defines the maximum amount of information required to perfectly describe a physical system up to the quantum level. Therefore, if a physical system is confined to finite energy within finite space, it must be described by a finite quantity of information \cite{Bekenstein:1980jp,Bekenstein:2004sh,Bekenstein:2000ai}. The Bekenstein bound is conceptually linked to the holographic principle \cite{tHooft:1993dmi,Bekenstein:1993dz,Fischler:1998st,Susskind:1998dq,Banks:2018aed}. The Bekenstein universal bound is expressed as:
\begin{eqnarray}
S \leq \frac{2 \pi~k_B R~ E}{\hbar~ c} \label{Bekenstein}
\end{eqnarray}
where $S$ is the thermodynamic entropy, $R$ is the radius of a sphere that encloses the physical system, and $E$ is the  energy of the physical system. Defining the information $H$ entropy as $S= 2 \pi k_B H$  that gives the number of bits contained in the quantum states in the sphere that encloses the physical system, we get:
\begin{eqnarray}
H \leq \frac{ R~ E}{\hbar~ c}
\end{eqnarray}
Earlier studies have delved into the relationship between the Bekenstein universal inequality and the uncertainty principle, as mentioned in \cite{Bousso:2004kp,Buoninfante:2020guu}. However, these investigations overlooked the essential influence of entropy on the commutation relationship between position and momentum. By integrating the notion of entropy, the correlation between the Bekenstein bound and the Generalized Uncertainty Principle (GUP) becomes clear. We aim to provide a more detailed conceptual framework for understanding the equivalence between the Bekenstein bound and the bounds predicted by GUP. To this end, let's examine the common characteristics of both bounds.

\begin{enumerate}[label=(\roman*)]
    
\item Both  bounds are  universal \cite{Das:2008kaa,Ali:2011fa,Bekenstein:1980jp,Ali:2022ckm}.

 \item GUP bound implies maximum measurable momentum at which the physical system is completely described as we show in section (\ref{sec:LGUP}). Bekenstein universal bound implies the maximum amount of information necessary to describe the physical system.  Both bounds aim for a complete and perfect description of physical reality.
    
    \item The bound implied by GUP varies from one physical system to another according to large number of studies  \cite{Das:2008kaa,Ali:2011fa,Pikovski:2011zk,Brau:1999uv,Casadio:2020rsj,Prasetyo:2022uaa,Acquaviva:2022yiq,Chevalier:2021xyw,Bosso:2018ckz,Feng:2016tyt,Das:2011tq,Quesne:2009vc}. This  means that the GUP bound depends on the geometric boundaries and mass of the physical system. The Bekenstein universal bound depends on the energy, the radius of the sphere that encloses the physical system, and the energy of the system. Both universal bounds vary from one physical system to another.
\item Bekenstein universal bound that can be  written as
\begin{eqnarray} \label{BBH}
H\leq \frac{\hbar^{\prime}}{\hbar} ~~~~~\text{where}~~~ \hbar^{\prime}= \frac{R~E}{c}   
\end{eqnarray}
This means that the Bekenstein universal bound introduces the entropy as it relates to an effective variation of the Planck constant given by $\hbar^{\prime}$.
On another side, The linear GUP model can be  written as 
\begin{eqnarray}
\label{gup}
[x_i, p_j] = i \hbar~\hspace{-0.5ex} \hspace{-0.5ex} \left[  \delta_{ij}\hspace{-0.5ex}
- \hspace{-0.5ex} \alpha\hspace{-0.5ex}  \left( p \delta_{ij} +
\frac{p_i p_j}{p} \right)
\right]=i~ \hbar^{\prime} \delta_{ij}\end{eqnarray}
where $\hbar^{\prime}=\hbar \left[ 1- \frac{4}{3} \alpha~p\right]$. The Bekenstein universal bound and the linear GUP introduce the same  physical concept of the effective variation of the Planck constant.
\end{enumerate}
Based on Eq. (\ref{BBH}) and Eq. (\ref{gup}), we obtain the following:
\begin{eqnarray}
\alpha_0 p=\frac{3}{4}\sqrt{ \frac{\hbar~c^3}{~ G}}\left(1-\frac{R~E}{\hbar~c}\right) \label{nature}
\end{eqnarray}
If we consider $E= M c^2$, where $M$ is the relativistic mass, four  cases of equation (\ref{nature}) follow:
\begin{itemize} [label=$\triangle$]
    \item When $M~R~c < \hbar$,  the value $\alpha_0~p$ is found to be positive, but according to linear GUP, $\alpha_0$ must be positive, so the momentum must be positive. This case matches particles such as Higgs, Z-Boson, W-Boson, kaon, pion,  and electron  \cite{Ali:2022ckm}. This case is only applicable for short lengths. 
    
    \item When  $M~R~c > \hbar$,  the value $\alpha_0~p$ is found to be negative, but according to linear GUP, $\alpha_0$ must be positive, so the momentum must be negative. The massive objects that match this case are all matters that are composed of protons up to the whole universe based on our computations in \cite{Ali:2022ckm} .i.e, long range.

    \item When $M~R~c = \hbar$, the value $\alpha_0~p$ is found to be zero, but according to linear GUP, $\alpha_0$ must be positive, so the momentum must be zero. This case corresponds to an object with Planck mass ($M=M_{p}$) and Planck length ($R=\ell_p$). This state is static and possibly explains why nature constants must be constants \cite{Bambi:2022lhq}. It is also beyond interaction because of its absolute zero momentum which may point to solid dark matter \cite{Bucher:1998mh}. According to our computation, no particle has been found that represents that state. The closest particles for that state are pions with ( $M~R~c=0.449 \hbar$), kaons ($M~R~c=1.458 \hbar$), Higgs ($M~R~c=0.4\hbar$), Z-Boson ($M~R~c=0.6 \hbar$) \cite{Ali:2022ckm}. This state forms a static bridge between the positive momentum case and the negative momentum case.  
    
    \item The maximum value of momentum $p$ happens  when $M=0$, which characterizes light. This means that the simultaneous reality between position and momentum happens when moving at the speed of light. Therefore it is a physical state  in nature. The wavefunction collapses in this state. In the next subsection, we compute the length scale of wavefunction collapse for electron and proton.

\end{itemize}
\subsection{Wavefunction collapse}
\par\noindent
The vanishing commutation relation between position and momentum  means that the wavefunction collapses. Let us consider two examples, the electron, and the proton. 
\subsubsection{Electron case}

\noindent
First, we compute the length at which the electron wavefunction collapses. We compute the linear GUP  bound for the electron using equation (\ref{nature}). For the electron that moves around an atom with an average speed $2.188\times 10^6$  \cite{Bohr:1913zba} and with considering the classical radius of electron and the mass of electron \cite{tiesinga2021codata} and substituting this in equation ((\ref{nature})), we find that the  GUP parameter is given by
\begin{eqnarray}
 \alpha_0 = 2.437\times 10^{24}.
\end{eqnarray}
When we consider this value in equation (\ref{minimallength}) to compute the length at which the wavefunction collapses (vanishing of commutation relation), we get:
\begin{eqnarray}
 \frac{4}{3}~\alpha_0 ~ \ell_p \approx 5.224  \times 10^{-11} ~~~\text{meter}.
\end{eqnarray}
The value $ 5.224  \times 10^{-11}$ meter is equal to, with a very good precision, the known value of Hydrogen atom radius given by $5.29  \times 10^{-11}$ \text{meter} \cite{tiesinga2021codata}. This means linear GUP could explain the radius value of a Hydrogen atom as the length at which electron wavefunction collapses. 

\subsubsection{Proton case}
\noindent
To make sure about our argument, we repeat the same computations for a proton moving inside a Hydrogen nucleus with an average speed ($0.2~c$) with considering proton mass and proton charge radius \cite{tiesinga2021codata}, we found that:
\begin{eqnarray}
 \alpha_0= 1.439\times 10^{20},
\end{eqnarray}
which gives minimum measurable length or the length at which proton wavefunction collapses according to the equation (\ref{minimallength}) as:
\begin{eqnarray}
  \frac{4}{3}~\alpha_0 ~ \ell_p \approx 10^{-15} \text{meter},
\end{eqnarray}
which agrees  with the measured Hydrogen nucleus radius \cite{tiesinga2021codata}. Once again, the radius of the hydrogen nucleus can be elucidated by linear GUP. This methodology can be readily applied to other particles and elements in nature. Linear GUP connects the quantum realm to the geometrical and gravitational domains by demonstrating wave function collapse.

\subsection{Solution of cosmological constant problem}
\par\noindent
The relationship between GUP and the Bekenstein bound prompts us to consider its implications on a cosmological scale \cite{Ali:2022ulp,AEGMV}. Previous research has explored the cosmological and astrophysical consequences of the Bekenstein bound \cite{Banks:2018aed,Cai:2001ur,Smolin:1995nk}. To test the validity of our equation, we can examine whether it can offer an explanation for cosmological phenomena, such as the cosmological constant problem. In our study \cite{Ali:2022ulp,AEGMV}, we computed  the Bekenstein bound for the entire universe, as described in equation (\ref{nature}), to solve the cosmological constant problem. The solution can be summarized as follows. According to equation (\ref{nature}), the effective Planck constant for the universe reads:
\begin{eqnarray}
r_{\text{univ}}~m_{\text{univ}}~c = \hbar^{\prime}_{\text{univ}}~. \label{huniv}
\end{eqnarray}
Using the astrophysical measurements $r_{\text{univ}}=1.37 \times 10^{26}$~m and $m_{\text{univ}}=2\times 10^{52}$ kg \cite{beadnell1940mass,davies2008goldilocks,Gott:2003pf}, the effective Planck constant evaluates to
\begin{eqnarray}
\hbar^{\prime}_{\text{univ}}= 8.2\times 10^{86} ~\text{J$\cdot$s}=7.82\times 10^{120} \hbar  \label{heffuniv}~.
\end{eqnarray}
When we replace $\hbar$ with the effective Planck calculated in Eq. (\ref{heffuniv}) in the basic definition of vacuum energy density in QFT. In this sense, the vacuum energy density reads: 
\begin{eqnarray}
\rho^{\text{eff}}_{\text{vac}} c^2&\approx&
\frac{1}{(2\pi \hbar^{\prime}_{\text{univ}})^3}\int^{P^{\prime}_{Pl}}_0 d^3p ~~ (\frac{1}{2} \hbar^{\prime}_{\text{univ}} \omega_p) \label{Vaccumdensityeff} \\&=&\frac{1}{(2\pi \hbar^{\prime}_{\text{univ}})^3}\int^{P^{\prime}_{Pl}}_0 d^3p \frac{1}{2} (c~p)\nonumber\\
&=& \frac{c^7}{16 \pi^2 \hbar^{\prime}_{\text{univ}} G^2} =\frac{c^7}{16 \pi^2\times 2.514\times 10^{121} \hbar~G^2} \nonumber\\&=&3.76\times 10^{-10}~~ \text{Joules}~,\label{EQFT}
\end{eqnarray}
where we used the effective Planck momentum $P^{\prime}_{Pl}= \sqrt{\hbar^{\prime}_{\text{univ}}c^3/G}$ as the cutoff of the integration. Comparing the calculated value of the effective energy density   $\rho^{\text{eff}}_{\text{vac}}$ with the observed value $\rho^{\text{obsv}}_{\text{vac}}= 5.3\times 10^{-10} ~~\text{Joules}$ \cite{ParticleDataGroup:2020ssz}, we find \cite{Ali:2022ulp}:
\begin{eqnarray}
\frac{\rho^{\text{obsv}}_{\text{vac}}}{\rho^{\text{eff}}_{\text{vac}}}= 1.41~.
\label{ratio_rho)}
\end{eqnarray}
This resolves the problem of the cosmological constant and also reinforces the reliability of the connection between linear GUP and the Bekenstein universal bound as expressed in equation (\ref{nature}). It is also consistent quantum gravity models that considered the cosmological constant as an infrared boundary condition\cite{Banks:2018jqo}. The proposed solution does not rely on any free parameters or additional dimensions, in essence, \emph{the universe explains itself}. Our recently published research \cite{Ali:2022ckm,AEGMV} contains a more detailed conceptual explanation. \\
 A recent investigation proposed a conceptual association between spin and the Bekenstein universal bound \cite{Acquaviva:2020qbc}. We have recently identified a conceptual link between the linear GUP and spin \cite{Ali:2021oml}. In this letter, we argue that the linear GUP represents the Bekenstein universal bound within the realm of quantum mechanics. As a result, the previous study \cite{Acquaviva:2020qbc} is consistent and offers conceptual reinforcement for the relationship between the linear GUP, spin \cite{Ali:2009zq}, and the Bekenstein universal bound. The research in \cite{Iorio:2022ave} examines the fascinating implications of the GUP for symmetries.
\section{GUP and Covariant bound}

\begin{figure}[h]
    %\centering
    \includegraphics[scale= 0.4]{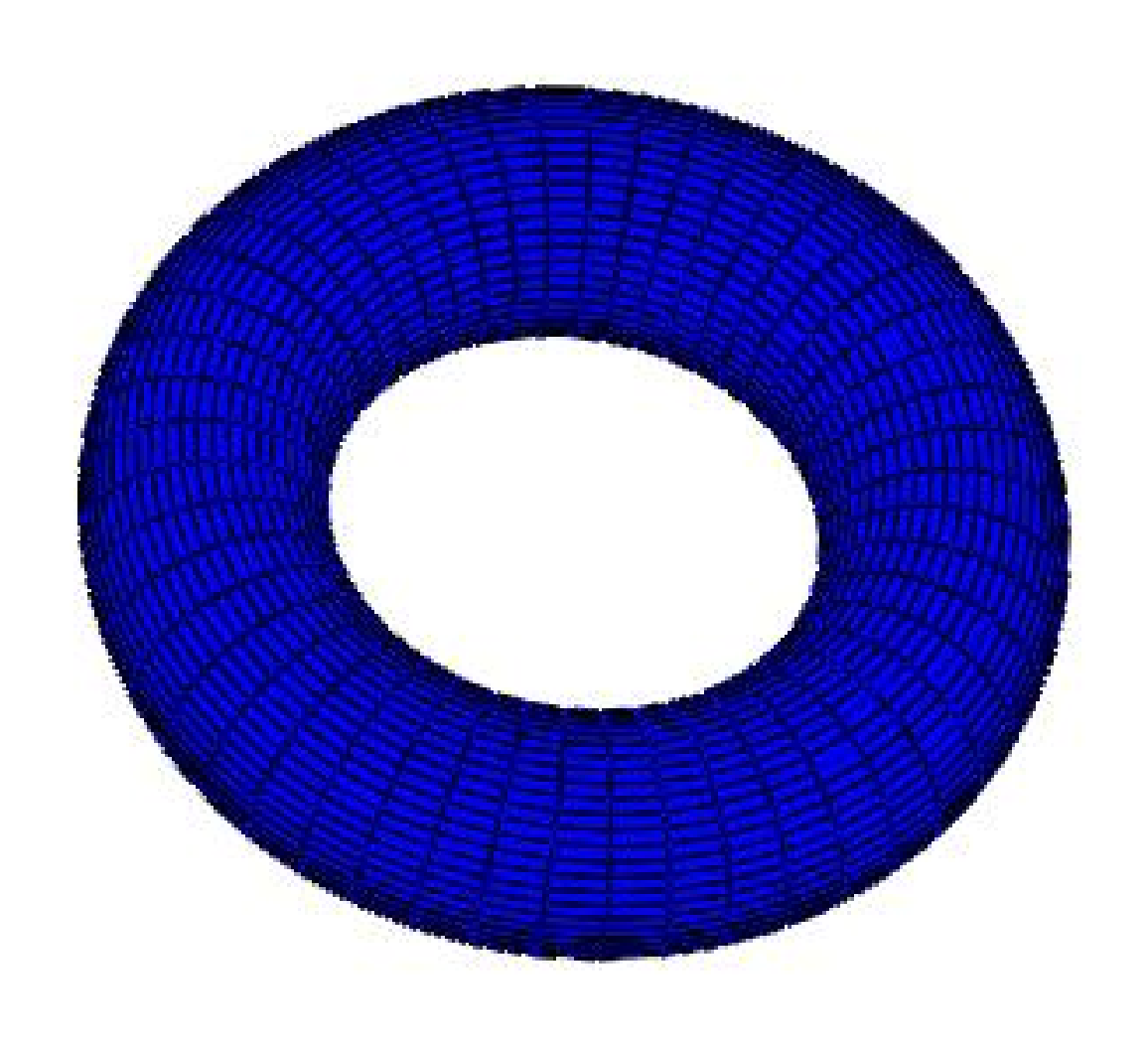}
    \caption{A 3-torus representing the physical system in the conext of GUP/minimal length models with $r_s$ as inner radius and $R$ as outer radius }
    \label{fig:enter-label}
\end{figure}

In previous sections, we have investigated the possible connection between the Bekenstein bound ($MRc$) and the GUP, which shed light on phenomena such as the wavefunction collapse and potential solutions to the cosmological constant problem. This leads us to question, what does the covariant version of the Bekenstein bound look like?   Before answering this question, let's revisit our recently published research \cite{Ali:2022ckm}, in which we have  noticed an interesting behavior in the value of $MRc$. When we examined particles, the value was found to oscillate around the Planck constant. But when we looked at nuclei, it varied within a few orders of magnitude of the Planck constant. This gave rise to the equation
\begin{eqnarray}
MRc= \alpha^{\prime} \hbar \label{universal}
\end{eqnarray}
where $\alpha^{\prime}$ is obtained from experimental measurements of the particle's mass and charge radius. For example, we have pions with ($MRc=0.449 \hbar$), kaons ($MRc=1.458 \hbar$), the Higgs boson ($MRc=0.4\hbar$), and the Z-Boson ($MRc=0.6 \hbar$) \cite{Ali:2022ckm}. We could express the equation (\ref{universal}) as:
\begin{eqnarray}
&& M~R~c= \alpha^{\prime} \hbar \label{shape} \implies \\&&  r_s~R= 2 \alpha^{\prime} \ell_{Pl}^2\label{shape1}
\end{eqnarray}
where $r_s$ is the Schwarzschild radius that correspond to the  mass of the physical system. The parameter $\alpha^{\prime}$  is an expression of entropy when comparing Eq. (\ref{shape}) with Bekenstein bound in Eq. (\ref{Bekenstein}). This comparison yields the following equation
\begin{equation}
  \frac{S}{k_B}\leq  2 \pi\alpha^{\prime}  \label{S-alpha}   
\end{equation}
When we substitute Eq. (\ref{S-alpha}) into Eq. (\ref{shape1}), we get the following relation
\begin{equation}
   \frac{S}{ k_B}  \leq \frac{ 4 \pi r_s~ R}{4\ell_{Pl}^2}
\end{equation}
Then the Bekenstein bound is represented by an area of 3-Torus  with inner radius $r_s$ and outer radius $R$ as
\begin{eqnarray}
  \frac{S}{k_B}\leq  \frac{A_{\text{3-Torus}}}{4 \pi \ell_{Pl}^2} \label{QGdiscovery}
\end{eqnarray}
where $A_{\text{3-Torus}}=4~\pi^2~r_s~ R $
The relation expressed in equation (\ref{QGdiscovery}) provides a geometric avenue for visualizing the GUP. In line with this conceptualization, we suggest that all physical systems might be modeled as a 3-Torus, with the inner radius equating to the Schwarzschild radius $r_s$ associated with the mass of the physical system, and the outer radius $R$ equal to the charge radius or in geometrical terms, the outer radius represents the radius of the smallest sphere that can enclose the physical system. This is illustrated in figure \ref{fig:enter-label}.

%The equation in \ref{shape} can be rewritten as
%\begin{eqnarray}
%    4 \pi^2 r_s R= 8 \pi^2 \alpha^{\prime} \ell_{Pl}^2
%\end{eqnarray}
%We get the area of the Torus as 
%\begin{eqnarray}
%    A= 4~\pi^2~r_s~R= 8~\pi^2~\alpha^{\prime} \ell_{Pl}^2= 4~S~\ell_{Pl}^2
%\end{eqnarray}
%where we used Bekensetin entropy area law. This implies that the entropy of the physical system visualized as a torus could be written as 
%\begin{eqnarray}
%    S= 2 \pi^2 \alpha^{\prime}
%\end{eqnarray}
\noindent
In this context, the 3-torus representation of any physical system may signify both a covariant and geometric form of the GUP, reinforcing its universality as evidenced in \cite{Das:2008kaa,Ali:2011fa}. Furthermore, our methodology is different from Bousso's covariant bound  \cite{Bousso:1999xy} which is defined in terms  the smallest surface area $A$ encloses the physical system. Yet, our suggestion goes a step further by visualizing this area in the form of a 3-Torus and linking this covariant bound with the GUP or minimal measurable length theories. The 3-torus forms an accurate geometric description of Bekenstein bound. GUP introduces a novel way to include entropy in the foundational rules of quantum mechanics. As indicated by \cite{CaboBizet:2022hpz}, entropy might be the foundational reason behind GUP. By considering Eq. (\ref{S-alpha}) and Eq. (\ref{QGdiscovery}), we can represent the GUP models as:
\begin{eqnarray}
    \Delta x \Delta p \geq \frac{S}{ k_B}~\frac{\hbar}{2} \\
    \Delta x \Delta p =\frac{A_{\text{3-Torus}}}{4 \pi^2 \ell_{Pl}^2}  \frac{\hbar}{2}
\end{eqnarray}
Here, the involvement of entropy sheds light on the principle of uncertainty. This suggests not only a potential bridge between quantum mechanics and thermodynamics at their core but also provides a deeper understanding of the role of entropy in shaping quantum behaviors.  With this perspective and the geometric connection of GUP with the area, we're excited to explore more about how GUP is connected to the entropy-area law in the next section.

\section{entropy-area law and  GUP}
\label{sec:EAL}
\par\noindent
In this section, we discuss the correspondence between  the proposed conjecture in equation (\ref{nature}) and Bekenstein-Hawking entropy-area law \cite{Bekenstein:1973ur,Hawking:1975vcx} 
\begin{eqnarray}
S_{\text{BH}}= \frac{c^3 k_B A_{\text{BH}}}{4 G \hbar} \label{BH}
\end{eqnarray}
Recently, entropy-area law was modified by a von Neumann entropy term in order to preserve the second law of thermodynamics, as well as to preserve information inside and outside the horizon. a coarse-grained entropy of the black hole was suggested in \cite{Penington:2019npb,Maldacena:2020ady,Almheiri:2019psf,wall2012proof,ryu2006holographic} that obeys the second law of thermodynamics, {\it i.e.}
\begin{eqnarray}
\label{SBH}
S_{\text{BH}}= \frac{ c^3 k_B A_{\text{BH}}}{4 G \hbar}+ S_{\text{matter}},
\end{eqnarray}
With considering $A_{\text{BH}}= 4 \pi R_s^2$ and $R_s=2~G~M/c^2$ of a black hole, we get:
\begin{eqnarray}
S_{\text{BH}}-S_{\text{matter}} =  2 \pi k_B \frac{M R_s c}{\hbar}
\end{eqnarray}
that can be written in terms of Information entropy as follows:
\begin{eqnarray}
    H_{\text{BH}}-H_{\text{matter}}= \frac{M R_s c}{\hbar}
\end{eqnarray}
Using equation (\ref{nature}), we get:
\begin{eqnarray}
    H_{\text{BH}}-H_{\text{matter}}= 1-2~\alpha~p
\end{eqnarray}
that can be rearranged as follows:
\begin{eqnarray}
    2~\alpha~p=\left(1+H_{\text{matter}}- H_{\text{BH}}\right) \label{free}
\end{eqnarray}
This suggests that the concept of momentum (motion) in physics may have an information-based origin \cite{Jacobson:1995ab, Wald:1993nt}. In essence, as Wheeler stated ``it from bit" \cite{wheeler2018information}. This could have implications for the holographic principle \cite{tHooft:1993dmi, Bekenstein:1993dz, Fischler:1998st, Susskind:1998dq, Banks:2003ta, Banks:2018aed, Bousso:1999xy, Casini:2008cr}. From this perspective, a complete portrayal of physical reality implied by linear GUP is conceptually equal to retaining information by including the von Neumann entropy correction in the entropy-area law. In other words, \emph{completeness is equivalent to information preservation}. Furthermore, equation \ref{free} provides additional insight into the four different scenarios discussed in section (\ref{sec:BUB}). This can be demonstrated as follows:

\begin{itemize} [label=$\nabla$]

\item  {\bf{Mixing states and aging of atoms}}: Positive momentum (i.e. $M R c<\hbar$) corresponds to the case ($1+H_{\text{matter}} \geq H_{\text{BH}}$). Positive momentum only exists for a short distance as we have shown in section (\ref{sec:BUB}).  Positive von Neumann entropy  quantifies the dissipative system. This case represents mixed states in quantum mechanics. This  may explain  how atoms age \cite{Raizen:2022xlv}. This case exists only at a short distance.\vspace{0.1cm} 
    
    \item   Negative momentum (i.e. $M R c~> \hbar$) corresponds to  the case ($1+H_{\text{matter}} < H_{\text{BH}}$). 
    
    %Negative momentum only exists at long distances as we have shown in section (\ref{sec:BUB}). This case can exist if von Neumann entropy is allowed to be negative. The negative von Neumann entropy was shown in \cite{Cerf:1995sa}  to correspond to  quantum entangled systems. Additionally, It was shown  in \cite{Cerf:1995sa} that  negative information  suggests a consistent interpretation of quantum informational processes with a unified description of classical correlations along with quantum entanglement.
    
    %The negative momentum case that we presented in section(\ref{sec:BUB}) can be used to explain the expansion of the universe and at the same time, it corresponds to negative von Neumann entropy that describes quantum entangled systems according to equation (\ref{free}). This sets an equivalency between the expansion of the universe and quantum entanglement, or in other words, the whole universe could be a quantum entangled system or quantum encoder \cite{Cotler:2022weg,Akal:2022qei}. Time is identified by expansion and based on our finding expansion is found to be a quantum entangled system, therefore time can be related to the complexity of the quantum entangled system \cite{Susskind:2014rva,Stanford:2014jda}. According to equation (\ref{free}),  the value of linear GUP parameter $\alpha_0$ quantifies the complexity for every physical object.  \vspace{0.1cm} 

    \item   {\bf{Certainty state }}: When $p$ reaches the maximum value ((i.e.$M=0$)), the ratio  $H_{\text{matter}}/{H_{\text{BH}}}=1$ that corresponds to a completeness state, certainty state or simultaneous reality   which is realized  by a physical object moving with the speed of light. This state  explains the wavefunction collapse  as we have shown in section (\ref{sec:BUB}).  \vspace{0.1cm} 
    \item  {\bf{Pure state and nature constants}}: When $p=0$ (i.e.$M R c~=M_p \ell_p c= \hbar$), we have  $H_{\text{BH}}= 1+H_{\text{matter}}$. 
\end{itemize}
\par\noindent
These four fundamental cases formulate the basic information description of nature. Nature  follows mathematical logic. We speculate that these four fundamental cases may  be assigned with Four-valued logic \cite{belnap1977useful}. We postulate  the correspondence to be as follows:
\par\noindent
\begin{table}[ht]
\centering
\begin{tabular}{|c|l|l|}
\hline
\textbf{Value} & \multicolumn{1}{c|}{\textbf{Stands For}} & \multicolumn{1}{c|}{\textbf{Physical State}} \\ \hline
1              & True                                      & $1 + H_{\text{matter}} \geq H_{\text{BH}}$   \\ \hline
0              & False                                     & $1 - H_{\text{matter}} < H_{\text{BH}}$      \\ \hline
Z              & Open circuit                              & $H_{\text{matter}} = H_{\text{BH}}$          \\ \hline
X              & Do not care                               & $1 + H_{\text{matter}} = H_{\text{BH}}$      \\ \hline
\end{tabular}
\caption{Description of Values and Their Corresponding Physical States}
\label{table:physical_states}
\end{table}

\par\noindent The deterministic nature of Bekenstein universal bound sheds light on the deterministic nature of physical reality. Several studies  followed the same understanding of nature such as \cite{tHooft:2020tuu,tHooft:1999imj,Page:1982fk,Jensen:1984gu,Donadi:2020aqz,Adler:1995iv,Hance:2021yik,Hossenfelder:2020wtj,Singh:2020gfn,Vrugt:2021sfu,Singh:2020jvy,Bohm:1951xw,Nelson:1966sp,Bassi:2012bg}. It is logical to conclude the following equivalency:
Determinism $\equiv$ Preserving information $\equiv$. This equivalency means that the universe  follows a feedback loop  or cycle. Linear GUP implies a cyclic universe \cite{Salah:2016kre} that could match Penrose's perspective  of conformal cyclic cosmology \cite{penrose2010cycles}.

\section{$SO(4)$ and $SO(1,3)$} \label{sec:sym}
\par\noindent
A question that naturally arises is: what does a negative momentum scenario signify, particularly in relation to spacetime? To address this question, we examine the underlying symmetry group of spacetime for these four fundamental scenarios. These four fundamental cases fully describe nature. Therefore, a four-dimensional space is required to fully account for them. To simplify, we assume that the four-dimensional space is Euclidean and is defined by its line element:

\begin{eqnarray}
dr^2= dx_1^2+ dx_2^2+dx_3^2+dx_4^2 \label{line}
\end{eqnarray}
where $x_1,x_2,x_3,x_4$ point to the four dimensions. We define the velocity in this space as 
\begin{eqnarray}
v_{i}=\frac{dx_i}{dx_1} ,~~~~~\text{where}~~~i=2,3,4
\end{eqnarray}
where $x_1$  plays the role of time in this four-dimensional space. Let us start with a positive momentum state. In that state, $x_i$ and $x_1$ must be real values (i.e. $x_1=c~t$) to give positive momentum values. This means time is like a real spatial dimension. Therefore, the line element reads
\begin{eqnarray}
 dr^2= c^2~dt^2+ dx_2^2+dx_3^2+dx_4^2 \label{timeless}
\end{eqnarray}
which is a real four-dimensional Euclidean space. In that state, time behaves similarly to a spatial dimension, implying that the symmetry group fully describing the state is the $SO(4)$ group, a real compact Lie group with six generators that is also known as the rotation group in four-dimensional space. As previously explained, positive momentum is limited to short-range distances, and hence, the positive momentum state is fully characterized by a compact $SO(4)$ group. This case illustrates that time behaves like spatial dimensions at short distances, and as a result, it can be reversed. Experimental evidence has demonstrated that quantum correlations can reverse the thermodynamic arrow of time at short distances, which is in agreement with our interpretation \cite{micadei2019reversing, PhysRevLett.109.064501,hofmann2017heat}.Let us consider the negative momentum case. To understand the spacetime origin of negative momentum. The momentum $p^2=p_i p_i$. Momentum can only be negative only if $p_i$ is imaginary. The momentum is the mass multiplied by the velocity. Mass can not be negative or imaginary, but velocity can be imaginary if the time is allowed to be like  an imaginary spatial dimension in that state (i.e $x_1=i~c~t$). In that state, the line element in equation (\ref{line}) reads
\begin{eqnarray}
dr^2= -c^2~dt^2+ dx_2^2+dx_3^2+dx_4^2 \label{lorentz}
\end{eqnarray}
which is the Lorentz spacetime. Therefore the negative momentum state is fully described by a Lorentz group $SO(1,3)$. It is known that $SO(1,3)$ is a non-compact group. Therefore, linear GUP implies $SO(4)$ at a short distance and $SO(1,3)$ once the wavefunction collapses at a relatively long distance. $SO(4)$ is only valid at a short distance, which is consistent with studies that found  QCD Hamiltonian is mapped onto an SO(4) representation which describes the QCD spectra at low energies with full preservation of flavor symmetry 
\cite{PhysRev.171.1691, Ramirez-Soto:2021jyh,Yepez-Martinez:2017jxj, Yepez-Martinez:2017ajq,Yepez-Martinez:2016xpb}. Additionally, it is  consistent with  studies on  finding that $SO(4)$ is the symmetry group of Hydrogen atom\cite{PhysRevA.77.034102,Szriftgiser:2020ybh,lieber19684,perelomov1966lorentz}. Additionally, it has been found that the solution to the SU(2) Yang-Mills field equations is invariant under SO(4) group  \cite{Belavin:1975fg,luscher1977so} which may be connected with instantons. In that sense, linear GUP implies that time reverses its role from being pure spatial dimensions (i.e. $SO(4)$) to be imaginary spatial dimension (i.e. $SO(1,3)$). This resonates with black holes and cosmological studies that suggest time has real and imaginary parts \cite{gibbons1977action,boyle2022thermodynamic}.

\section{Conclusions}\label{sec:con}
\par\noindent
In conclusion, the linear GUP suggests a simultaneous reality between position and momentum, which addresses the EPR argument. We hypothesize that there is a conceptual link between the linear GUP and Bekenstein's universal bound, a necessary bound for accurately describing a physical system. Consequently, the linear GUP may be a theory that integrates Bekenstein's universal bound within quantum mechanics, contributing to its completeness. This supports the notion that the minimal length is essential for physics \cite{Bosso:2022vlz}. Our conjecture also establishes a conceptual equivalence between the linear GUP and the von Neumann entropy correction term in the entropy-area law. This introduces an informational interpretation of the momentum (motion) concept. As a result, four fundamental cases in nature emerge. The first case may describe mixed states in quantum mechanics and the aging of atoms. The second case may explain the nature of time. The third case explains the wavefunction collapse, with the computed electron wavefunction collapse length scale being equal to $5.22\times10^{-11}$ meter, which agrees with the radius of a hydrogen atom ($5.29 \times 10^{-11}$ meter). Moreover, the hydrogen nucleus radius can be explained by the proton wavefunction collapse. In this sense, the linear GUP connects the quantum world with the geometric/gravity world through its ability to demonstrate the wavefunction collapse. Recent studies have found that gravity/geometry could be emergent from quantum \cite{Suh:2021mpp, Suh:2022kds, Das:2022tbv, Seiberg:2006wf}. A simultaneous reality implied by the linear GUP may be related to a timeless state \cite{Ali:2021ela}. The fourth case explains why nature's constants are constant as a result of zero momentum. The deterministic nature of the Bekenstein universal bound implies that determinism is equivalent to preserving information. This equivalence may suggest that the universe is a feedback loop/cycle, consistent with the cyclic universe predicted by the linear GUP and potentially aligning with Penrose's insight into conformal cyclic cosmology \cite{penrose2010cycles}. In a recent collaboration \cite{ERminimal}, we developed an Einstein-Rosen (ER) bridge using string T-duality corrected black holes, and established a connection with the Generalized Uncertainty Principle (GUP). This  geometric framework could potentially deepen our understanding of quantum entanglement for particle/antiparticle pairs. Our results showed the feasibility of an ER bridge, where the horizon area aligns with the Bekenstein minimal area bound in extreme mass configurations, possibly bearing implications for gravitational self-completeness and quantum mechanical mass limits. Given the observed associations between GUP, EPR, and ER, we concluded that GUP may function as a model representing the ER=EPR conjecture \cite{Maldacena:2013xja}.

\bibliographystyle{apsrev4-1}
\bibliography{ref.bib}{}

\end{document}